\documentclass[prd,twocolumn,tightenlines,showpacs,nofootinbib,superscriptaddress,nobalancelastpage]{revtex4-1} 

\usepackage{graphics}     
\usepackage{graphicx}     
\usepackage{longtable}     
\usepackage{url}          
\usepackage{bm}           
\usepackage{natbib}
\usepackage{textpos}
\usepackage{amsfonts,amsmath,amssymb,mathrsfs,bm,amsthm}
\usepackage{float}
\restylefloat{table}
\usepackage{booktabs}
\usepackage{array}
\usepackage{tabu}
\usepackage{dcolumn}
\usepackage{rotating}
\usepackage{ulem}

\usepackage{amsmath}
\usepackage{amssymb}
\usepackage{graphicx}
\usepackage{epstopdf}
\usepackage{inputenc}

\newcommand{\RN}[1]{%
  \textup{\uppercase\expandafter{\romannumeral#1}}%
}
\usepackage{soul} 
\usepackage{nicefrac}
\usepackage{amsthm}
\usepackage{color}
\usepackage{cancel}
\usepackage{appendix}
\usepackage{makecell}

\usepackage[colorlinks=true,linkcolor=black,citecolor=blue,urlcolor=blue,bookmarksopen]{hyperref}

\newcommand{\appropto}{\mathrel{\vcenter{
  \offinterlineskip\halign{\hfil$##$\cr
    \propto\cr\noalign{\kern2pt}\sim\cr\noalign{\kern-2pt}}}}}
\renewcommand{\v}[1]{\boldsymbol{#1}}		

\begin{document}
\title{Time-reversal invariance violation in neutron-nucleus scattering}

\date{\today}

\author{Pavel Fadeev}
\thanks{pavelfadeev1@gmail.com}
\affiliation{Helmholtz Institute Mainz, Johannes Gutenberg University, 55099 Mainz, Germany} 

\author{Victor V.~Flambaum}
\thanks{v.flambaum@unsw.edu.au}
\affiliation{Helmholtz Institute Mainz, Johannes Gutenberg University, 55099 Mainz, Germany}
\affiliation{School of Physics, University of New South Wales, Sydney, New South Wales 2052, Australia}

\begin{abstract}
Parity (P) and time-reversal (T) violating effects are enhanced a million times in neutron reactions near p-wave nuclear compound resonances. Planning and interpretation of corresponding experiments require values
of the matrix elements of the T,P-violating nuclear forces between nuclear compound states. We calculate the
root-mean-square values and the ratio of the matrix elements of the T,P-violating and P-violating interactions
using statistical theory based on the properties of chaotic compound states. We present the results in terms of
the fundamental parameters in five different forms: in terms of the constants of the contact nuclear interaction,
meson exchange constants, QCD $\theta$-term, quark chromo-electric dipole moments $\tilde{d}_u$ and $\tilde{d}_d$, and axion interaction
constants. Using current limits on these parameters, we obtain upper bounds on the ratio of the matrix elements
and on the ratio of T,P-violating and P-violating parts of the neutron reaction cross sections. Our results confirm
that the expected sensitivity in neutron-reactions experiments may be sufficient to improve the limits on the
T,P-violating interactions.
\end{abstract}

\maketitle
\section{Introduction}
\label{Sec:Intro}

A very popular way to search for time-reversal (T) and parity (P) violation and to test unification theories is based on the measurements of electric dipole moments (EDMs) of elementary particles and atomic systems. So far this method
has produced stringent limits on EDMs which exclude or bound many models (see reviews in Refs.  \cite{Ginges2004,Pospelov2005,Engel2017,Yamanaka2017,Chupp2019}).
Studies of T,P-violating (also known as T,P-odd) effects via EDM also give limits on the axion and relaxion interactions~\cite{Stadnik2017}.
An efficient alternative method is measurement of T,P-odd effects in neutron-nucleus scattering. This method is motivated by the millionfold enhancement of parity violation in neutron reactions near $p$-wave nuclear compound resonances, which was predicted in Refs.~\cite{Sushkov80,Sushkov82,Flambaum84,Flambaum85}.
The first confirmation was obtained in experiments performed at the Joint Institute for
Nuclear Research in Dubna \cite{Alfimenkov1983,Alfimenkov1984}; then a very extensive experimental study was done in several laboratories, including the Joint Institute for Nuclear Research (Dubna), Petersburg Institute of Nuclear Physics, KEK (Tsukuba), and especially in Los Alamos (see reviews in Refs.~\cite{MitchellReview99,MitchellReview2001}).
This activity
continues now (see, for example, the recent experimental
paper \cite{Okudaira2018} and references therein). A similar mechanism of enhancement should work for the T,P-odd effects \cite{Bunakov1983,TRIV87,Gudkov1992,Flambaum_Gribakin95}.
An unusual statistics of P-violating and T,P-violating effects,
namely random-sign observables not vanishing upon
averaging, was demonstrated in Refs.~\cite{FlamGrib94,BerengutFlamGrib}. Experiments
searching for T,P-violating effects are in progress in Japan and
the United States~\cite{Snow2017,Okudaira2018,Palos2018, Kitaguchi2018, Beda2007,Bowman96}.

Without any enhancement, the effects of P violation in
low-energy nuclear reactions are extremely small, $\sim 10^{-7}$ (e.g., in the proton scattering on hydrogen and helium, and
neutron radiative capture by protons) \cite{Gericke2011}.
The formula for a P-violation effect near a $p$-wave compound resonance may be presented as~\cite{Sushkov80,Sushkov82,Flambaum84,Flambaum85} \footnote{We omit the numerical coefficient which depends on the specific
process induced by the neutron capture.}
\begin{align}\label{P}
    P \sim \frac{W_{sp}}{E_s-E_p} \sqrt{\frac{\Gamma_s^n}{\Gamma_p^n}}\,,
\end{align}
where $W_{sp}$ is the matrix element of the parity-violating interaction
mixing $s$ and $p$ resonances, $E_s-E_p$ is the energy interval between these resonances, and $\Gamma_s^n, \Gamma_p^n$ are the neutron
widths of these resonances.
We see that there are two reasons
for the enhancement of P violation near $p$-wave compound
resonances. First, in a nucleus excited by neutron capture the
interval $E_s-E_p$ between the chaotic compound states (resonances)
of opposite parity is very small, and this enhances by
three orders of magnitude the mixing of these states by the
weak P-violating interaction between nucleons. The second
reason is that the admixture of opposite-parity states allows
neutron capture in the $s$ wave to contribute to the $p$-wave
resonance. At small neutron energies the $s$-wave amplitude is
three orders of magnitude larger than the p-wave amplitude
($\sqrt{\Gamma_s^n /\Gamma_p^n} \sim 10^3$). As a result of these two $10^3$ factors, the
P-violating parts reach 1--10\% of reactions’ cross sections
and become accessible to experimental scrutiny. T,P-violating
effects are also produced by the parity-violating interaction;
therefore, Eq.~(\ref{P}) and the enhancement mechanism works for
them too \cite{Bunakov1983,TRIV87,Gudkov1992,Flambaum_Gribakin95}.

For the experiments to produce useful results we need
theory for their interpretation. At first glance, it seems impossible, since chaotic compound states are very complicated.
However, chaos allows us to develop a statistical theory,
similar to the Maxwell-Boltzmann theory for macroscopic
systems, which actually gives very accurate predictions. We
developed such a theory, including a method to calculate
matrix elements between chaotic states in finite systems (in
excited nuclei, atoms, and molecules)~\cite{Flam93,Flam93PRL,Flambaum1994V,Flambaum1995a,Flambaum1995b,FlamGrib2000,Bowman96V}. We briefly
present the ideas below.

An increase in the excitation energy of a nucleus increases
the number of its active particles $k$ and available orbitals $p$, leads to an exponential increase of the density of energy
levels $\sim p!/[(p-k)!k!]$, and brings the system into a state
where the residual interaction between particles exceeds the
intervals between the energy levels. The eigenstates $\left. \middle|  n \right\rangle= \sum_i C_i^n \left. \middle|  i \right\rangle$ become chaotic superpositions of thousands or even
millions of Hartree-Fock basic states $\left. \middle|  i \right\rangle$. All medium and heavy nuclei and atoms with an open $f$ shell have chaotic excited compound states in the discrete spectrum and/or
chaotic compound resonances. The idea of Refs.~\cite{Flam93,Flam93PRL} is to treat the expansion coefficients $C_i^n$ as Gaussian random variables, with the average values $\overline {C_i^n}=0$ and variance
\begin{equation}\label{C2}
\overline{(C_i^n)^2}=\frac{1}{\bar{N}} \Delta(\Gamma_{spr},E^n - E_i) \, ,
\end{equation}
\begin{equation}\label{Delta}
\Delta(\Gamma_{spr},E^n - E_i)=\frac{\Gamma_{spr}^2/4}{(E^n - E_i)^2 + \Gamma_{spr}^2/4} \, \, ,
\end{equation}
where $\bar{N} = \pi \Gamma_{spr} / 2d$ is the normalization constant found
from $\sum_i (C_i^n)^2 =1$, $d$ is the average energy distance between
the compound states (resonances) with the same angular momentum and parity, and $\Gamma_{spr}$ is the spreading width of the
component calculated using the Fermi golden rule \cite{ABohr};
$\bar{N}$ is called the number of principal components.\footnote{
Basis states $\left. \middle|  i \right\rangle$ with shell-model energies $E_i$ close to the energy
of a compound state $E^n$ (within the spreading width $\Gamma_{spr}$) have the highest weight ($\sim 1/\bar{N}$) and dominate in the normalization sum $\sum_i (C_i^n)^2 =1$. The number of such  states is $\bar{N}$.}  

We have tested this distribution of $C_i^n$ by the numerical
calculations of chaotic compound states in cerium and protactinium
atoms~\cite{Flambaum1994,FlambaumGribakina1995,FlambaumGribakina1996,FlambaumGribakina1998E,FlambaumGribakina1998A,FlambaumGribakina1999,Viatkina}, in highly charged ions with an open $f$ shell~\cite{GribakinGribakina1999b,FlambaumGribakina2002,DzubaFlambaum2002,DzubaFlambaum2013,Berengut,Harabati}, in the two-body random interaction model~\cite{Izrailev1996,Izrailev1997a,Izrailev1997b,Izrailev2000} and using an analytical approach~\cite{FlamGrib2000,IzrailevGribakin}.

The function $\overline{(C_i^n)^2}= \Delta(\Gamma_{spr},E^n - E_i) / \bar{N}$ gives the probability to find the basis component $\left. \middle|  i \right\rangle$ in the compound state $\left. \middle|  n \right\rangle$;  i.e. it plays the role of the statistical partition function.
The difference from the conventional statistical theory is that
the partition function depends on the total energy of the
isolated system $E^n$ instead of on the temperature of a system in
a thermostat [recall the Boltzmann factor $\exp{(-E_i/T)}$]. One
may compare this with the microcanonical distribution where the equipartition is assumed within the shell of the states with fixed energy $E_i$.

Expectation values of matrix elements of any operator $W$ in a chaotic compound state are found as $\overline{\left. \middle| \left\langle n \middle| \right. W \left. \middle|  n \right\rangle \middle| \right.^2}=\sum_{i} \overline{(C_i^n)^2} \left.  \middle| \left \langle i \middle| \right. W \left. \middle|  i \right\rangle \middle| \right.^2$. For example, this formula with $W=a^+_k a_k$  (the occupation-number operator) gives the distribution of the orbital occupation numbers in finite chaotic systems which replaces the Fermi-Dirac (or Bose-Einstein) distribution.\footnote{However, numerical calculations \cite{Flambaum1994, GribakinGribakina1999b,FlambaumGribakina2002,Izrailev1996} give occupation numbers which are close to the Fermi-Dirac distribution.}
 
Average values of the non-diagonal matrix elements
are equal to zero, $\overline{ \left\langle n \middle| \right. W \left. \middle|  m \right\rangle}=0$, while the average values of the squared matrix elements
$W^2 \equiv \overline{\left. \middle| \left\langle n \middle| \right. W \left. \middle|  m \right\rangle \middle| \right.^2}=\sum_{i,j} \overline{(C_i^n)^2} \overline{(C_j^m)^2} \left. \middle| \left\langle i \middle| \right. W \left. \middle|  j \right\rangle \middle| \right.^2$ 
are reduced to the sum of matrix elements between the Hartree-Fock basis states $ \left. \middle| \left\langle i \middle| \right. W \left. \middle|  j \right\rangle \middle| \right.^2$, where $W$ is any perturbation operator. The distribution of the matrix elements $\left\langle n \middle| \right. W \left. \middle|  m \right\rangle$  is Gaussian with the variance given by the $W^2$.
  
For the correlator between two different operators (e.g., P-violating
and T,P-violating) we obtain  $\overline{\left\langle n \middle| \right. W_P \left. \middle|  m \right\rangle \left\langle m \middle| \right. W_{T,P} \left. \middle|  n \right\rangle}=\sum_{i,j} \overline{(C_i^n)^2} \overline{(C_j^m)^2} \left\langle i \middle| \right. W_P \left. \middle|  j \right\rangle\left\langle j \middle| \right. W_{T,P} \left. \middle|  i \right\rangle$ \cite{Flam93,Flam93PRL,Flambaum1994V,Flambaum1995a, Flambaum1995b}. Note that our
theory predicts the results averaged over several compound resonances.
  
We have done many tests comparing the statistical theory results for electromagnetic amplitudes \cite{FlambaumGribakina1998A}, electron recombination rates \cite{GribakinGribakina1999b,FlambaumGribakina2002,DzubaFlambaum2002,DzubaFlambaum2013,FlambaumKozlov2015,Berengut,Harabati} and parity-violation effects \cite{Flam93,Flam93PRL} with the experimental data and with numerical simulations.
For example, we obtained a thousandfold enhancement of
the electron recombination rate with many highly charged
tungsten ions due to the very dense spectrum of chaotic
compound resonances \cite{FlambaumGribakina2002,DzubaFlambaum2002,DzubaFlambaum2013,FlambaumKozlov2015,Berengut,Harabati}. These results agree with all available experimental data and predict recombination rates for ions with a high ionization degree, where experiments are limited by existing techniques. Our results are important for thermonuclear reactors which are made from tungsten. Tungsten ions contaminate plasma and significantly affect the energy output.
 
Using the theory of chaotic nuclear compound resonances,
we calculate in this paper the ratio $w/v$ of the root-mean-square
values of the matrix elements of the T,P-odd ($w$) and P-odd ($v$) matrix elements. 
We show the results in terms of the
fundamental parameters in five different forms: in terms of the
constants of the contact nuclear interaction, meson exchange
constants, QCD $\theta$-term,  quark chromo-EDMs $\tilde{d}_u$ and $\tilde{d}_d$, and axion interaction constants. Using latest bounds on $\theta, \tilde{d}_u$ and $\tilde{d}_d$, and axion interaction constants we arrive at bounds on
the magnitude of possible T violation. In the Conclusion section
the results are compared with the expected experimental
sensitivity to the T,P-violating effects.

\section{P- and T,P-violating interactions}
\label{Sec:Ratio}

The ratio of the time-reversal-invariance violating (TRIV)
and parity violating (PV) parts of the neutron nuclear cross sections induced by mixing of $s$- and $p$-wave nuclear compound resonances, $\Delta \sigma_{PT} / \Delta \sigma_{P}$,  can be expressed as \cite{Gudkov2018,Gudkov1990,Okudaira2018}: 
\begin{equation} 
 \label{cross}
 \frac{\Delta \sigma_{PT}}{\Delta \sigma_{P}} = 
 \kappa \frac{\left\langle \psi_p \middle| W_{PT} \middle| \psi_s \right\rangle}{\left\langle \psi_p \middle| W_P \middle| \psi_s \right\rangle} \, .
\end{equation}
Here the factor $\kappa$ includes amplitudes of the partial neutron
widths which depend on spin channels $J = I \pm 1/2$, where $I$ is the spin of the target nucleus and $J$ is the spin of the compound resonance. For example, for $J=0$, one obtains $\kappa=1$, as in this case $\kappa$ does not depend on neutron partial
widths \cite{Gudkov1992,Bunakov1983,TRIV87}. 

The ratio  $\Delta \sigma_{PT} / \Delta \sigma_{P}$ for the neutron-deuterium scattering
was calculated in Ref.~\cite{Song2011}. However, experiments are planned for heavier nuclei where we expect a \mbox{millionfold} enhancement of the T,P-odd and P-odd effects.
 
In the short-range interaction limit, the PV operator $W_P$ and TRIV operator $W_{PT}$ are 
\begin{equation}  \label{Wp}
 W_P = \frac{G g}{ 2 \sqrt{2} m} \left\{ (\sigma {\bf p}) , \rho \right\} \, ,
\end{equation}

\begin{equation}  \label{Wpt}
 W_{PT} = \frac{G \eta}{ 2 \sqrt{2} m}  (\sigma \nabla  ) \rho \, .
\end{equation}
Here $G$ is the weak-interaction Fermi constant, $m$ is the nucleon mass, ${\bf p}$ and $\sigma$ are nucleon momentum and spin respectively, and $\rho$ is the nucleon density. Nucleon dimensionless constants $g_{p,n}$ and $\eta_{p,n}$  characterize the strength of the interactions. Note that in the standard definition of angular wavefunctions the matrix element of $W_P$ between bound states is imaginary (since the momentum operator ${\bf p}=-i \nabla$) and the matrix element of TRIV operator $W_{PT}$ is real.

We define $v^2$ to be the average of the absolute value of the squared PV matrix element, and $w^2$ to be the average value of the squared TRIV matrix element between the $s$ and $p$ compound resonances, such that
\begin{equation}
v = \sqrt{
\overline{
\left\langle \psi_p \middle| W_P \middle| \psi_s \right\rangle
\left\langle \psi_s \middle| W_P \middle| \psi_p \right\rangle
}
} \, ,
\end{equation}
\begin{equation}
w = \sqrt{
\overline{
\left\langle \psi_p \middle| W_{PT} \middle| \psi_s \right\rangle
\left\langle \psi_s \middle| W_{PT} \middle| \psi_p \right\rangle
}
} \, .
\end{equation}

Correlations might exist between the matrix elements of PV and TRIV interactions. The quantity parametrizing such correlations, the correlator, is defined as
\begin{align}
C &= 
\frac{
|\overline{
\left\langle \psi_p \middle| W_{P} \middle| \psi_s \right\rangle
\left\langle \psi_s \middle| W_{PT} \middle| \psi_p \right\rangle
}|
}{v \, w} \, .
\end{align}
The correlator, which takes values between zero and one, can be useful to deduce the values and signs of TRIV effects, since
much is already known about the PV effects. The correlator $C$ was calculated by the same technique as the mean-square
matrix element and was found to be \cite{Flambaum1995b}:
\begin{align}
\mid  C \mid &\approx 0.1 \, .
\end{align}
This result tells us that the correlations between the matrix elements are relatively small so we may neglect them.

\subsection{Rough estimate of \textit{w$/$v}}
Naively one would expect from Eqs.~(\ref{Wp}) and~(\ref{Wpt}) the following relation: $w / v \sim \eta / g$.
However, this ratio is actually $A^{1/3}$ times  smaller than the ratio of interaction constants \cite{Flambaum1995a}, where $A$ is the number of nucleons. Indeed, for $\nabla \rho$ in Eq.~(\ref{Wpt}),
\begin{equation} \label{nabrho}
\nabla \rho \sim \frac{\rho}{R_N} \sim \frac{\rho}{r_0 A^{1/3}}\, ,
\end{equation}
where $r_0$ is the internucleon distance, and $R_N=r_0 A^{1/3}$ is the nuclear radius.
The momentum in Eq.~(\ref{Wp}) is approximated as $p \sim p_F \sim \hbar/r_0$.
Thus, the ratio of matrix elements is smaller
than the ratio of interaction constants in Eqs.~(\ref{Wp}) and~(\ref{Wpt}) by a factor of $A^{1/3}$:
\begin{equation} 
 \frac{w}{v}  \sim \frac{\eta}{g A^{1/3}}  \, .
\end{equation}
For elements with the number of nucleons in the range 100--250, $A^{1/3} \approx 5$. A detailed discussion of this suppression factor
including many-body effects can be found in Ref.~\cite{Flambaum1995a}.
 \subsection{Dependence of matrix elements on nucleon interaction constants}
\begin{figure}
\includegraphics[width=0.45\textwidth]
{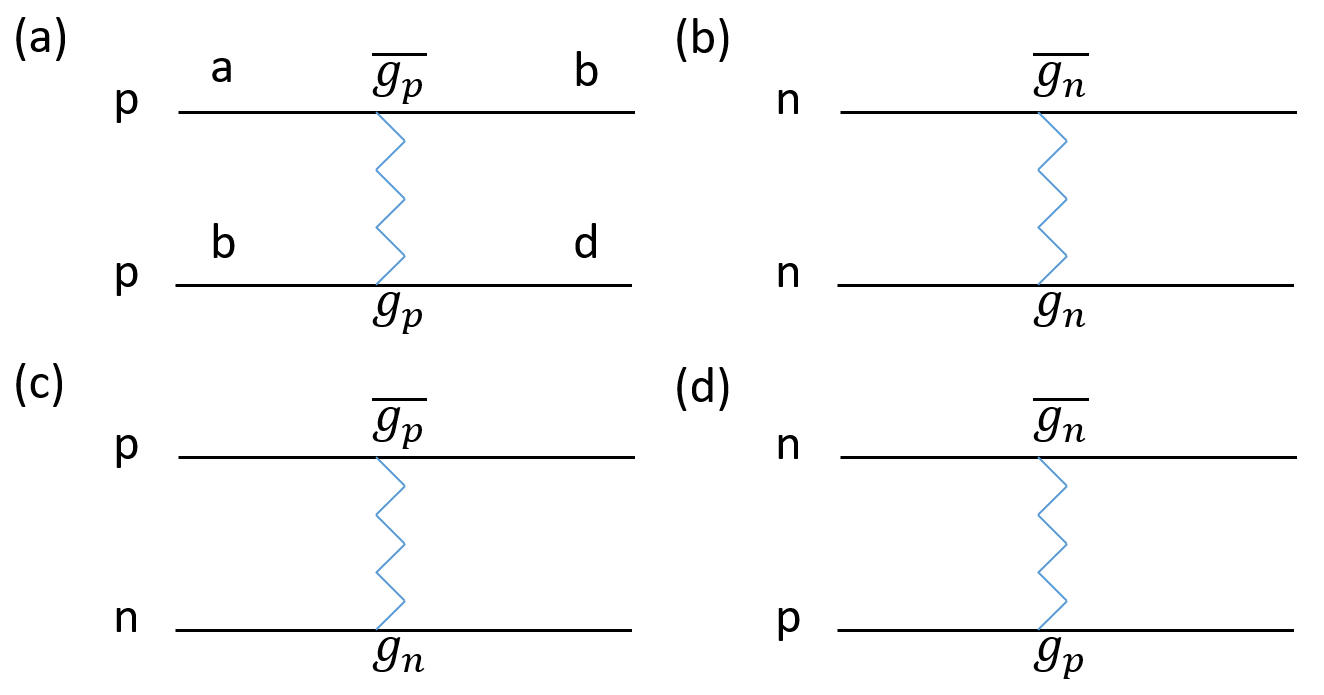}
\caption{Possible configurations of weak interactions $V_{ab,cd}$ \cite{Flam93PRL} within the nucleus between protons (p) and neutrons (n). In each diagram, the upper vertex is P-violating. Constants $g_p$ and $g_n$ characterize
the strength of the interactions. (a) Interactions between two protons; (b) interactions between two neutrons; (c) and (d) interactions between protons and neutrons,
which contribute to the squared PV matrix element $v^2$ by $(V_{np} + V_{pn}) ^2 = V^2_{np}+ V^2_{pn}+2 V_{np}V_{pn}$. When summing matrix elements in (c) and (d), the terms $V_{np}V_{pn}$ have random signs and the result is much smaller than the sums of $V^2_{np}$ and $V^2_{pn}$.} 
\label{fig:Interactions}
\end{figure}

A general expression for the root-mean-square value of the
matrix element $v$ the PV  operator (and the matrix element $w$ of the TRIV operator) was derived in Ref.~\cite{Flam93PRL}:
\begin{widetext}
\begin{align}\label{Vnunu}
    v = \frac{1}{\sqrt{\Bar{N}}}
    \left\{ 
    \sum_{abcd} \,
    \nu_a
    \left( 1 - \nu_b \right)
    \nu_c
    \left( 1 - \nu_d  \right)
    \frac{1}{4} \left.
    \middle|
    V_{ab,cd} - V_{ad,cb} 
    \middle| \right. ^2
    \Delta \left( \Gamma_{spr}, \epsilon_a - \epsilon_b + \epsilon_c - \epsilon_d \right)
    \right\}^{\frac{1}{2}} \, .
\end{align}
\end{widetext}
Here $\nu$ are the orbital occupation numbers given by the Fermi-Dirac distribution in an excited nucleus, numerical values of the matrix elements of the two-nucleon interaction $V_{ab,cd}$ (see Fig.~\ref{fig:Interactions}) are presented in Refs.~\cite{Flam93PRL,Flambaum1995b}, and $\Delta \left( \Gamma_{spr}, \epsilon_a - \epsilon_b + \epsilon_c - \epsilon_d \right)$ is the ``spread" $\delta$ function [Eq.~(\ref{Delta})]  of the change in energy $\epsilon_a - \epsilon_b + \epsilon_c - \epsilon_d$. Equation~(\ref{Vnunu})
has a clear interpretation. The $\Delta$ function means in fact an approximate energy conservation with an accuracy up to the spreading width $\Gamma_{spr}$ (since the single-particle states
are not stationary states in this problem). In the case $\Gamma_{spr} \to 0$ we have  $\Delta \left( \Gamma_{spr}, \epsilon_a - \epsilon_b + \epsilon_c - \epsilon_d \right) \to (\pi\Gamma_{spr}/2)\delta(\epsilon_a - \epsilon_b + \epsilon_c - \epsilon_d) $. To have a transition, initial states must be occupied (this gives  $\nu_a$ and  $\nu_c$) and final states empty  (this gives  $1- \nu_b$ and  $1- \nu_d$). 
 
The dependence of $v$ and $w$ on the nucleon interaction constants $g$ and $\eta$ [which appear in Eqs.~(\ref{Wp}) and (\ref{Wpt})] can be presented in the following form  \cite{Flam93PRL,Flambaum1995b}:
    \begin{equation} \label{vvv}
        v = \frac{1}{\sqrt{\Bar{N}}} \sqrt{
    \left( \Sigma_{pp}^{(P)} g_p \right)^2 
    +
    \left( \Sigma_{nn}^{(P)} g_n \right)^2 
    +
    \left( \Sigma_{pn}^{(P)} g_p g_n \right)
    } \, ,
\end{equation}
\begin{equation} \label{www}
    w = \frac{1}{\sqrt{\Bar{N}}} \sqrt{
    \left( \Sigma_{pp}^{(PT)} \eta_p \right)^2 
    +
    \left( \Sigma_{nn}^{(PT)} \eta_n \right)^2 
    +
    \left( \Sigma_{pn}^{(PT)} \eta_p \eta_n \right)
    } \, ,
    \end{equation}
where
$g_p$ and $g_n$ are proton and neutron weak constants --- they characterize the strength of the P-odd weak potential; $\eta_p, \eta_n$ are constants that characterize the strength of the T,P-odd potential, and $\Sigma$ are sums of the weighted squared matrix
elements of the weak interaction between nucleon orbitals defined in Eq.~(\ref{Vnunu}). Contributions of the cross terms $\Sigma_{pn}^{(P)} g_p g_n$ and  $ \Sigma_{pn}^{(PT)} \eta_p \eta_n$ are small compared to the other terms since
they contain products of different matrix elements which have random signs, while in the terms containing squared interaction constants all contributions are positive. 

Therefore, we can present $v$ and $w$ in the following form:
 \begin{align} \label{vgngp}
     v = K_P \sqrt{g_n^2 + k g_p^2} \, ,
 \end{align}
  \begin{align} \label{wgngp}
     w = K_{PT} \sqrt{ \eta_n^2 + k \eta_p^2} \, .
 \end{align}
 The coefficient $k$ should be slightly smaller than 1 since in heavy nuclei the number of neutrons $N= 1.5 Z$, where $Z$ is the number of protons.
  To make a simple estimate of the sensitivity of $v$ and $w$ to changes in the interaction constants, we assume in the next step that $\Sigma$ from Eqs.~(\ref{vvv}) and (\ref{www}) are proportional to the number of interaction terms in the nucleus.    
 There are $Z^2/2$ interaction terms between protons, $N^2/2$ such terms between neutrons, and $ZN$ terms between a proton and a neutron (Fig.~\ref{fig:Interactions}). Thus we can write
\begin{align}
    k = \frac{Z^2+ 2ZN}{N^2 +2ZN} = 0.76 \, .
\end{align}

Numerical calculations of $v$ and $w$ have been done in Refs.~\cite{Flam93PRL,Flambaum1994V,Flambaum1995a, Flambaum1995b} for specific values of the interaction constants $g_p, g_n, \eta_p$ and $\eta_n$. The values of these constants have been updated since those calculations. Therefore, we would like to find updated values of these constants to insert into Eqs.~(\ref{vgngp}) and (\ref{wgngp}). The general expressions for $g_p$ and $g_n$ are~\cite{Flam97,FKS84,Flam93,Desplanques80}
\begin{align} \nonumber \label{g_p}
g_p &= 2\times 10^5 \, W_\rho \Bigg[
176 \frac{W_\pi}{W_\rho} f_\pi -19.5 h^0_\rho -4.7 h^1_\rho +1.3h^2_\rho \\  
&  -11.3 (h^0_\omega + h^1_\omega) \Bigg] \, , \\
 \nonumber \label{g_n}
g_n &= 2\times 10^5 \, W_\rho \Bigg[
-118 \frac{W_\pi}{W_\rho} f_\pi -18.9 h^0_\rho +8.4 h^1_\rho -1.3h^2_\rho\\  
& -12.8 (h^0_\omega + h^1_\omega) \Bigg] \, ,
\end{align}
where $f$ and $h$ are the weak $NN$-meson couplings, and $W_\rho$ and $W_\pi$ are constants which account for the repulsion between nucleons at small distances as well as for the finite range of the interaction potential. We take $W_\rho=0.4$ and $W_\pi=0.16$ as in Refs. \cite{Flam97,FKS84}.

For the choice of constants $g_p = 4 , g_n=1$ \cite{Noguera86}, numerical calculations give \cite{Flam93PRL}
     \begin{align} \label{WP_old}
    v =& K_P \sqrt{1 + 16 k }
        = 2.08 \, \text{meV} \, .
     \end{align}    
We calculate updated values for $g_p$ and $g_n$, using the best values of the constants $h$ from Desplanques, Donoghue, and
Holstein (DDH) \cite{Desplanques80} with an updated  $f_\pi \equiv h^1_\pi$, which was recently derived by lattice QCD methods  \cite{Wasem2012,Haxton2013}. Such calculations give  $g_p = 2.6 \, , g_n=1.5$ (Table~\ref{tab1}). Using these values with Eq.~(\ref{WP_old}), we have 
          \begin{align} \label{Pnew} 
  v_{updated} &= 
 2.08 \, \text{meV}  \frac{ \sqrt{1.5^2 +2.6^2 k }}{ \sqrt{1 + 16 k }}
  =  1.56 \, \text{meV} \, ,
     \end{align}
where in the last step we used $k=0.76$.\footnote{For $k=1$ we would get $1.51$ meV. Thus we see that our result is not very sensitive to the value of $k$.}
This theoretical estimate is in excellent agreement with the experimental value $1.39^{+0.55}_{-0.38} \, \, \text{meV}$ \cite{Bowman90, Frankle91}.

Numerical calculations were done for $\eta_p = \eta_n$ and gave  $w =0.2 | \eta_n | $ meV \cite{Flambaum1995b}. Using this result and  Eq.~(\ref{wgngp}) we obtain
    \begin{align} \label{PT3}  w_{updated} =& 0.15 \, \text{meV} \sqrt{ \eta_n^2 + 0.76 \eta_p^2}  \, .
    \end{align}
\begin{table}[H]
\caption{Values of $g_p$ and $g_n$ based on the meson exchange constants
from different publications (left-hand column): 
Desplanques,
Donoghue, and Holstein (DDH) \cite{Flam97,Desplanques80}; Noguera and Desplanques
(ND) \cite{Noguera86}; Dubovik and Zenkin (DZ) \cite{DZ1986}; Feldman, Crawford,
Dubach, and Holstein (FCDH) \cite{FCDH1991}.
In the line of Wasem \cite{Wasem2012} we use the best DDH values for all the values of $h$ except $f_\pi = h^1_\pi$, which was recently derived by the lattice QCD methods  \cite{Wasem2012,Haxton2013} to be $h^1_\pi = 1.1 \cdot 10^{-7}$.
Recent experiment measuring P-violation in the neutron radiative capture by proton \cite{npdgamma}
gave $h_{\pi}^{1}=[2.6 \pm 1.2 (stat.) \pm 0.2 (sys.)] \times 10^{-7}$ which is larger than the theoretical estimate $h^1_\pi = 1.1 \cdot 10^{-7}$. Using this experimental value, and the rest from DDH, gives slightly larger $g_p=3.4 \pm 0.8$ and smaller $g_n=0.9 \pm 0.6$ which are close to the values $g_p=4$ and $g_n=1$ used in the numerical calculation of P-violation in Ref. \cite{Flam93PRL}.  Corresponding value  of the matrix element $v=1.8 \pm 0.4$ meV is consistent with both Eq. (\ref{WP_old}) and Eq. (\ref{Pnew}).}
\centeringֻֻ
\begin{tabular}{l|*{3}{c|}r|}
  Reference & ~$g_p$~ &~$g_n$~ \\ \hline 
 ~DDH (1980) \cite{Flam97,Desplanques80} ~ & $4.5$ & $0.2$   \\[0.2cm] 
  ~ND (1986) \cite{Flam93PRL,Noguera86}~ & $4$ & $1$   \\[0.2cm]
  ~DZ (1986) \cite{DZ1986} ~ & $2.4$ & $1.1$   \\[0.2cm]
  ~FCDH (1991) \cite{FCDH1991}~ & $2.7$ & $-0.1$   \\[0.2cm]
    ~Wasem (2012) \cite{Wasem2012}~ & $2.6$ & $1.5$   \\[0.2cm] 
      ~NPDGamma (2018) \cite{npdgamma}~ & $3.4$ & $0.9$   \\ \hline 
\hline 
\end{tabular}
\label{tab1}
\end{table} 
 
 \subsection{The ratio $w/v$ expressed via meson exchange constants, QCD $\theta$-term, quark chromo-EDMs $\tilde{d}_u$ and $\tilde{d}_d$, and axion exchange constants}
 Now we can express the ratio $w/v$ in five different ways: as a function of $\eta_p, \eta_n$, by $\pi_0$ meson-exchange coupling constants with the nuclei, by QCD CP-violation parameter $\theta$,  by quark chromo-EDMs, and finally by axion exchange constants.
 
First, to express the ratio $w/v$ as a function of $\eta$, we use Eqs.~(\ref{Pnew}) and (\ref{PT3}) to obtain
     \begin{align} \label{ratio1}
         \frac{w}{v} &=
         0.10  \sqrt{
 \eta_n^2 
    +
     0.76 \eta_p^2  
    } \, . 
     \end{align}
If, following Refs.~\cite{Flambaum1995b,Flambaum2014}, we take $|\eta_p| = |\eta_n|$, the ratio in Eq.~(\ref{ratio1}) becomes   
       \begin{align} \label{ratio2}
         \frac{w}{v} &=
 0.13  |\eta_n|  \, . 
     \end{align}   
     
Second, the T,P-odd nuclear forces are dominated by  $\pi_0$ meson exchange. Such an exchange is described by the interaction \cite{Haxton83,Khriplovich2000,Dmitriev03}
\begin{align} \label{piTP} \nonumber
\mathcal{W} \left(\v{r}_1 - \v{r}_2 \right) &= - \frac{\bar{g}}{8 \pi m_N} 
\left[ \nabla_1 \left(
\frac{e^{-m_{\pi}r_{12}}}{r_{12}}
 \right) \right] \cdot
 \left\{ \left( \v{\sigma}_1 - \v{\sigma}_2 \right) \right.
 \\  \nonumber
 &  \times \left. 
\left[
\bar{g}_0 \v{\tau}_1 \cdot \v{\tau}_2
+ \bar{g}_2 \left( \v{\tau}_1 \cdot \v{\tau}_2 - 3 \tau_{1z} \tau_{2z} \right)
 \right] 
  \right.
 \\  
 &\left. +
 \bar{g}_1 \left(
 \tau_{1z} \v{\sigma}_1
 - \tau_{2z} \v{\sigma}_2
  \right)
  \right\} \, ,
\end{align}
where $\bar{g}=13.6$ is the strong-force T,P-conserving $\pi NN$ coupling constant, $\bar{g}_0, \bar{g}_1$, and $\bar{g}_2$, are the strengths of the isoscalar, isovector, and isotensor T,P-violating couplings, respectively, $m_N$ is the nucleon mass, $m_{\pi}$ is the pion mass, $\v{\sigma}$ is the nucleon spin, $\v{\tau}$ is the nucleon Pauli isospin matrix in vector form, and $r_{12}$ is the separation between nucleons.
The coupling constants $\eta$ can be expressed in terms of $\bar{g}$ \cite{Flambaum2014}:
     \begin{align}
     -\eta_p= \eta_n =5 \times 10^{6} \bar{g} \left( \bar{g}_1 +0.4 \bar{g}_2 - 0.2\bar{g}_0 \right) \, .
      \end{align}
Then we have
     \begin{align}\label{wvg}
        \frac{w}{v} &= 0.13 |\eta_n| 
        = |6.5 \times 10^5 \bar{g} \left( \bar{g}_1 +0.4 \bar{g}_2 - 0.2\bar{g}_0 \right)| \, . 
     \end{align}
     
Third, using the previous results $\bar{g}\bar{g}_0 = -0.37 \theta$ \cite{Crewther1980} ,where $\theta$ is the QCD CP-violation parameter, and $\bar{g}\bar{g}_1=\bar{g}\bar{g}_2=0$, we can write the ratio $w/v$ as a function of $\theta$:
     \begin{align}
        \frac{w}{v} &
        = 4.8 \times 10^4 |\theta| \, .
     \end{align}
    Using updated results \cite{Yamanaka2017,Vries2015}
    \begin{align}
       \bar{g}\bar{g}_0 &= -0.2108 \, \theta \, , \\
       \bar{g}\bar{g}_1 &= 46.24 \times 10^{-3} \theta \, ,
    \end{align}
    we can write, still with $\bar{g}\bar{g}_2=0 $,
         \begin{align} \nonumber
        \frac{w}{v} &=  
        5.7 \times 10^4
          |\theta| \, .
        \end{align}
    Using the current limit on $\theta$, obtained from constraints on neutron EDM, $|\theta| <  10^{-10}$ \cite{Yamanaka2017}, we obtain
\begin{equation} \label{limittheta}
     w/v <  10^{-5}  \, .
\end{equation} 

Fourth, we can connect our result to the quark chromo-EDM $\tilde{d}$ \cite{Pospelov2005}:
    \begin{align}
        \bar{g}\bar{g}_1 &= 4 \times 10^{15} \left( \tilde{d}_u - \tilde{d}_d \right) / \text{cm} \, , \\
        \bar{g}\bar{g}_0 &= 0.8 \times 10^{15} \left( \tilde{d}_u + \tilde{d}_d \right) / \text{cm} \, . 
    \end{align}
    Then
         \begin{align}
\frac{w}{v} 
&= | \left. 6.5 \times 10^{20} \left( 
 4 \left( \tilde{d}_u - \tilde{d}_d \right) 
  -0.16  \left( \tilde{d}_u + \tilde{d}_d \right) 
  \right) | \right.  / \text{cm} \, .
     \end{align} 
    Using the current limits (Table IV in \cite{Swallows2013}; see also Ref.~\cite{Graner2016})
\begin{align}
|\tilde{d}_u - \tilde{d}_d |&< 6 \times 10^{-27} \text{cm} \, ,
\\  
|\frac{1}{2}\tilde{d}_u + \tilde{d}_d| &< 3 \times 10^{-26} \text{cm} \, ,
\end{align}    
 we obtain
    \begin{equation} \label{limitd}
    w/v < 2 \times 10^{-5} \, .
    \end{equation}

Finally,  a T,P-violating interaction, similar to the pion-exchange-induced Eq. (\ref{piTP}), may be due to exchange by any
scalar particle which has both scalar (with the interaction constant $g^s$) and pseudoscalar  (with the interaction constant $g^p$) couplings to nucleons. The most popular examples are the dark-matter candidates axion \cite{Moody,Marsh} and relaxion \cite{Graham,Gupta,Flacke}, which have  very small masses.\footnote{The limits on the T,P-violating electron-nucleon interactions mediated by the axion or relaxion exchange from EDM measurements
were obtained in Ref. \cite{Stadnik2017}, where more references may be found.} A numerical estimate shows that due to the long range of the interaction the matrix elements in the small-mass case ($e^{- mr} \approx 1)$ are $\sim 1.5$ times larger than the pion exchange matrix elements; i.e., we have instead of Eq. (\ref{wvg}) the following estimate:
\begin{align}\label{wva}
        \frac{w}{v} & \sim  |1 \times 10^6 g^s g^p|   \, . 
     \end{align}
The limit on $g^sg^p$ may be obtained from the proton EDM calculation,\footnote{The calculation is similar to that for electron EDM \cite{Stadnik2017}.}
\begin{equation} \label{dp}
d_p=\frac{g^sg^pe}{8 \pi^2 m_p}\,,
    \end{equation}
and measurement \cite{Graner2016}, $|d_p|<2\times 10^{-25}e$ cm, $|g^s g^p| < 1 \times 10^{-9}$. Using limits from the proton EDM and the $^{199}$Hg nuclear-Schiff-moment measurements in Ref. \cite{Swallows2013}, the authors of Ref. \cite{musolf} concluded that the limit on $|g^s g^p|$ is between $10^{-9}$ and $10^{-11}$. This gives a rather weak limit on $w/v$ induced by axion exchange:    
\begin{equation} \label{limita}
    w/v < 10^{-3} - 10^{-5} \, .
\end{equation}
\section{Conclusion}
\label{Sec:Discussion}
Using the bound in Eq.~(\ref{limitd}), and assuming $\kappa \approx 1$ in Eq~(\ref{cross}) (which is reasonable \cite{Gudkov2018} and matches experimental results \cite{Okudaira2018,Palos2018}), we arrive at
\begin{align}
 \frac{\Delta \sigma_{PT}}{\Delta \sigma_{P}} \underset{\sim}{<} 2 \times 10^{-5}
 \, .
\end{align}
The limit based on the axion exchange in Eq.~(\ref{limita}) is weaker. The vurrent expected experimental sensitivity is
\cite{Gudkov2014,Palos2018} 
\begin{equation} 
 \frac{\Delta \sigma_{PT}}{\Delta \sigma_{P}}_{\text{exp. sensitivity}} < 10^{-6}
 \, .
\end{equation}
Thus we confirm that the expected experimental sensitivity in neutron reactions may be sufficient to improve the limits on the TRIV interactions, or possibly to detect them. 
\section*{Acknowledgements‏}
We thank Anne Fabricant for editing the manuscript. We are grateful to William Michael Snow for informing
us about the new measurement of the constant $h_\pi^1$ \cite{npdgamma}. This work is supported by the Australian Research Council and Gutenberg Fellowship.

\bibliographystyle{prsty}

\end{document}